\documentclass[twocolumn, aps, pra, amsmath, amssymb, nofootinbib, superscriptaddress, longbibliography, floatfix, table-of-contents, eqsecnum,10pt]{revtex4-2}

\usepackage[]{graphicx}
\usepackage{mathrsfs}

\usepackage[colorlinks = true, linkcolor=teal, citecolor=teal, urlcolor  = teal]{hyperref}
\usepackage{xurl}

\usepackage{array}
\usepackage{comment}
\usepackage{mathtools}
\usepackage{times}
\usepackage{type1cm}
\usepackage{lettrine}
\usepackage[english]{babel}
\usepackage{microtype}
\usepackage{booktabs}
\usepackage[boxed, vlined]{algorithm2e}
\usepackage[margin=0pt, font=small, labelfont=bf, labelsep=endash, justification=centerlast, labelsep=colon]{caption}
\usepackage{braket}
\usepackage[toc,page]{appendix}

\usepackage{amssymb}
\usepackage{mathrsfs}
\usepackage{subfig}
\usepackage{multirow}
\usepackage{etoolbox}
\usepackage{breqn}
\usepackage{float}
\usepackage{nicematrix}
\usepackage{makecell}
\usepackage{tikz}

\newtheorem{theorem}{Theorem}

\makeatletter
\preto\maketitle{%
  \begingroup\lccode`~=`,
  \lowercase{\endgroup
  \let\saved@breqn@active@comma~
  \let~}\active@comma 
}
\appto\maketitle{%
  \begingroup\lccode`~=`,
  \lowercase{\endgroup
  \let~}\saved@breqn@active@comma 
}

\makeatother

\begin{document}

\title{Rigorous no-go theorems for heralded linear-optical state generation tasks}

\author{Deepesh Singh }
\thanks{These two authors contributed equally}
\affiliation{Centre for Quantum Computation and Communication Technology, School of Mathematics and Physics,
The University of Queensland, Brisbane, Queensland 4072, Australia}

\author{Ryan J. Marshman}
\thanks{These two authors contributed equally}
\affiliation{Centre for Quantum Computation and Communication Technology, School of Mathematics and Physics,
The University of Queensland, Brisbane, Queensland 4072, Australia}

\author{Luis Villegas-Aguilar}
\affiliation{Queensland Quantum and Advanced Technologies Research Institute, Centre for Quantum Computation and Communication Technology, Griffith University, Yuggera Country, Brisbane, Queensland 4111, Australia}
\affiliation{Center for Macroscopic Quantum States (bigQ), Department of Physics, Technical University of Denmark, 2800 Kgs. Lyngby, Denmark}

\author{Jens Eisert}
\affiliation{Dahlem Center for Complex Quantum Systems, Freie Universität Berlin, 14195 Berlin, Germany}

\author{Nora Tischler}
\email[]{n.tischler@griffith.edu.au}
\affiliation{Queensland Quantum and Advanced Technologies Research Institute, Centre for Quantum Computation and Communication Technology, Griffith University, Yuggera Country, Brisbane, Queensland 4111, Australia}

\begin{abstract}
A major challenge in photonic quantum technologies is developing strategies to prepare suitable discrete-variable quantum states using simple input states, linear optics, and auxiliary photon measurements to identify successful outcomes. Fundamentally, this challenge arises from the lack of strong non-linearities on the single-photon level, meaning that photonic state preparation based on linear optics cannot benefit from the deterministic gate-based approach available to other physical platforms. Instead, the preparation of quantum states can be probabilistically implemented using single photons, linear-optical networks, and photon detection. However, determining whether an input state can be transformed into a target state using a specific measurement pattern---a problem that can be mapped to deciding the feasibility of a system of polynomial equations---is a complex problem in general. To solve it, we apply the Nullstellensatz Linear Algebra algorithm from algebraic geometry to quantum state generation; this can provide definitive no-go results by proving infeasibility when the state preparation task in question has no solution. We demonstrate this capability to validate and establish lower bounds on the physical resource requirements for the realization of several ubiquitous optical states and gates. 
\end{abstract}

\maketitle

\newtheorem{lemma}{Lemma}
\newtheorem{sublemma}{Lemma}[lemma]

\newcommand{\overbar}[1]{\mkern 1.5mu\overline{\mkern-1.5mu#1\mkern-1.5mu}\mkern 1.5mu}

\section{Introduction}

Entangled photonic states are crucial resources in quantum computing, sensing, cryptography, and communication \cite{gisin_quantum_2007,obrien_photonic_2009,flamini_photonic_2018,pirandola_advances_2020,Slussarenko}. The deterministic generation 
of entangled states from separable states would 
require strong non-linearities, but implementing such non-linearities in practical optical platforms remains extremely challenging. By contrast, linear optics is experimentally accessible, but it permits only a limited set of deterministic transformations in 
the space of photonic quantum states for any reasonable encoding. Various approaches have been developed to formally characterize the limitations of linear optics for deterministic transformations \cite{PARELLADA2023107108,PhysRevA.89.062329,PhysRevA.92.053844}. However, arbitrary state transformations can be implemented probabilistically using a combination of single photons, linear optics, and photon detection, provided that one has access to sufficiently many photons \cite{KLM}. This way of effectively generating arbitrary state transformations is particularly relevant in the quest to create linear-optical quantum computers and simulators in a measurement-based fashion \cite{PhysRevLett.95.010501,PhysRevLett.96.020501,PhysRevA.74.042343,Fusion,QuantumPhotoThermodynamics,Xanadu}. In this work, we focus on fundamental limitations---i.e., no-go results---associated with probabilistic transformations in linear-optical systems when a limited number of photons are available.

The probabilistic nature of photonic state generation necessitates the identification of successful events. On the highest level, there are two general approaches: (1) post-selected schemes, whereby the successful events can be sifted from the larger data set by measuring the photons that make up the state of interest (the target state), and (2) heralded schemes, where only auxiliary photons in
ancillary modes are measured to provide an independent signal that the target state has been created. Unlike post-selection, where the target state is destroyed upon identification, heralding allows the quantum state to be utilized freely, without restrictions on subsequent interference of modes, and allows multiplexing to increase the probability of success \cite{forbes2025heraldedgenerationentanglementphotons}. This makes heralded state generation schemes far more useful, and as such, they will be the focus of this article.

The feasibility of a state generation task---determining whether an input state can be transformed into a target state using a given measurement pattern---is generally a difficult problem. It can be reformulated as deciding whether a corresponding system of polynomial equations possesses a solution. Ref.~\cite{PhysRevA.76.063808} has proposed the use of Gr\"{o}bner basis techniques from algebraic geometry to find solutions to state generation tasks, not entirely dissimilar to new approaches making use of algebraic geometry ideas for quantum error correction \cite{WillsAlgebraic}. This approach not only determines the feasibility of state generation tasks but also provides the solutions, i.e., suitable transformations. However, computing a Gr\"{o}bner basis can be computationally expensive, with worst-case time complexity being doubly exponential 
\cite{rolnick2019robust}. Furthermore, it is progress-free, so the full calculation must be completed to obtain any results.

The Gr\"{o}bner basis approach may be excessive for scenarios where only the feasibility of a state generation task needs to be determined, without necessarily finding a specific linear-optical circuit that enables its implementation. As an alternative, we propose using the 
\emph{Nullstellensatz Linear Algebra} (NulLA) algorithm to prove the infeasibility of quantum state generation tasks \cite{DELOERA20111260}. NulLA can be used as a progressive algorithm in the search for \emph{infeasibility certificates} of increasing degrees, corresponding to increasing computational hardness. If a certificate is found, it provides definitive proof that the system of polynomial equations has no solutions, thus attesting to the impossibility of generating the target state using a given input state and measurement scheme. This kind of rigorous proof stands in stark contrast to the typical approach, which is based on numerical search \cite{Fldzhyan2021, PhysRevA.102.012604, gubarev2021fockspaceperspectiveoptimal}. When such a search finds no solution, one may suspect but cannot know for sure that no solution exists. The crucial advantage provided by algebraic geometry approaches is the rigorous nature of the results obtained.

Since NulLA focuses solely on deciding the feasibility of the system without finding the solution itself (if one exists), we anticipate it to be particularly useful in such decision problems. These problems are directly linked to resource requirements. In particular, our approach allows proving that the generation of a given target state is impossible when starting with a certain number of single photons. Such lower bounds on the resource requirements can be used to prove the resource optimality of known solutions and provide important insights into problems without known solutions. 
Although we mainly focus on the problem of generating photonic quantum states from deterministic single-photon sources, the technique can also be extended from deterministic single-photon inputs to handle probabilistic photon-pair sources, as realized with non-linear processes such as spontaneous parametric down-conversion and four-wave mixing. Moreover, it can be extended from photonic quantum state generation to photonic quantum gates.

The rest of this article is structured as follows: 
Section~\ref{sec:prelim} explains the mapping of the heralded state generation task to solving a system of polynomial equations and proving its infeasibility using the NulLA algorithm. Section~\ref{sec:properties} outlines the scaling of the technique and discusses simplifications applicable to the input state and measurement pattern, which enable a categorization based solely on the number of available input photons. Section~\ref{sec:applications} highlights some example applications of the technique and Section \ref{sec:conclusion}, contains concluding remarks.

\section{Technique to prove infeasibility \label{sec:prelim}}

In this section, we revisit how a heralded photonic quantum state generation task can be cast as a system of polynomial equations, which was first described in Ref.~\cite{PhysRevA.76.063808}
(and used in Ref.\ \cite{PhysRevLett.106.013602}). We then propose to apply another technique from algebraic geometry, the NulLA algorithm \cite{DELOERA20111260}, for a targeted approach to deciding the feasibility of heralded photonic state preparation tasks.

\subsection{Formulation of a heralded photonic state generation task as a system of polynomial equations}

This subsection provides an overview of the steps presented in Ref.~\cite{PhysRevA.76.063808} to determine a linear transformation that probabilistically evolves an input state to a target state conditional on measuring a heralding pattern. The linear transformation is treated as unknown, while the input state, target state, and heralding pattern are assumed to be given. The linear transformation is not forced to be unitary in the first instance because (a) this choice makes the problem amenable to convenient algebraic geometry tools and (b) once determined, the transformation can subsequently be made unitary by rescaling and extending its dimensions through the inclusion of ancillary modes. 

In a nutshell, in the approach taken, the input state vector 
$\vert \psi_{ \mathrm{in}} \rangle$ is first transformed into $\vert \psi_{ \mathrm{out}} \rangle$ via the linear transformation represented by the matrix $A$. From $\vert \psi_{ \mathrm{out}} \rangle$, an output state vector $\vert \psi_{ \mathrm{post}} \rangle$ is created by conditioning on a particular measurement pattern in the heralding modes. For the photonic quantum state generation task, this output state, which will generally depend on the unknown transformation matrix $A$, needs to be equivalent---that is, equal up to a nonzero scaling factor---to the target state vector $\vert \psi_{ \mathrm{tar}} \rangle $. 
All multi-mode states with a fixed photon number can be written as homogeneous polynomials of creation operators, of degree equal to the total number of photons and with the number of variables equal to the number of modes in the corresponding 
state. The equivalence of any two states can then be reduced to the equivalence of the corresponding two polynomials of creation operators. Equating the corresponding monomial coefficients further leads to a system of polynomial equations where the variables are the elements of the unknown transformation matrix $A$. Whether this system of polynomial equations possesses a solution determines the feasibility of the state generation task, and a solution of the system of polynomial equations provides $A$, which in turn can be further processed to yield linear-optical circuits that implement the state generation.

In more detail, and following the convention in Ref.~\cite{PhysRevA.76.063808} for ease of cross-referencing, we take the number of modes in the linear transformation to be $N$, and the number of heralding modes to be $M$. The number of input photons in the $i^{\mathrm{th}}$ mode is $n_{i}$, with the total number of input photons being $n$, i.e., 
\begin{equation}
n \coloneqq \sum\limits^{N}_{i=1} n_{i} . 
\end{equation}
In the chosen heralding pattern, the number of measured photons in the $j^{\mathrm{th}}$ mode is given by $m_{j}, \forall (N-M+1) \leq j \leq N$, and the total number of heralding photons 
is 
\begin{equation}
m \coloneqq \sum\limits^{N}_{j=N-M+1} m_{j}.
\end{equation}
Therefore, the output state vector $\vert \psi_{ \mathrm{post}} \rangle$ that is produced conditioned on measuring the heralding signal is an $(N-M)$-mode, $(n-m)$-photon state.

The input state to the transformation is assumed to be a product state in Ref.~\cite{PhysRevA.76.063808} and hence its corresponding polynomial $R$ can be represented by a single monomial
\begin{align}
    \vert \psi_{ \mathrm{in}} \rangle &:= R(a^{\dagger}_{1,\mathrm{in}}, \dots , a^{\dagger}_{N,\mathrm{in}}) \vert 0 \rangle \nonumber \\
    &= \prod^{N}_{i=1
    } \frac{1}{\sqrt{n_{i}!}} a^{\dagger n_{i}}_{i, \mathrm{in}} \vert 0 \rangle.
\end{align}
By rescaling if necessary, one can ensure that a linear transformation $A$ has spectral norm $\Vert A \Vert \leq 1$. Then $A$ is unitary if all singular values equal one, and otherwise unitarity can be recovered later by embedding it in a larger transformation \cite{PhysRevA.76.063808, PhysRevX.8.021017}.
Thus, we can consider $A$ as a general linear transformation without assumptions on its singular values. The creation operators for the modes $1 \leq i \leq N$ evolve as $a^{\dagger}_{i,\textrm{in}} 
\mapsto
\sum\nolimits^{N}_{j=1} A_{i,j}a^{\dagger}_{j}$, where the subscript ``in" indicates input modes, while no subscript indicates output modes. 
The input state vector evolves as 
\begin{align}
        \vert \psi_{\mathrm{out}} \rangle & :=F(a^{\dagger}_{1}, \dots , a^{\dagger}_{N}) \vert 0 \rangle \nonumber \\
         &= \prod^{N}_{i=1}\frac{1}{\sqrt{n_{i}!}} \bigg(\sum\limits^{N}_{j=1} A_{i,j} a^{\dagger}_{j} \bigg)^{n_{i}} \vert 0 \rangle,
\end{align}    
where 
$F(a^{\dagger}_{1}, \dots , a^{\dagger}_{N})$ is a homogeneous polynomial of degree $n$ in the creation operators $a^{\dagger}_{i}$ for $1 \leq i \leq N$. The coefficients of the monomials in $F$ are themselves homogeneous polynomials of degree $n$ belonging to the polynomial ring $\mathbb{C}[A_{i,j}]$.

The effect of measuring the last $M$ modes of the linear network and heralding on the photon measurement pattern $(m_{1},\dots,m_{M})$ results in the 
heralded state vector
\begin{align}
        \vert \psi_{ \mathrm{post}} \rangle & := G(a^{\dagger}_{1}, \dots , a^{\dagger}_{N-M}) \vert 0 \rangle \nonumber \\
         =& \langle m_{1},m_{2},\dots, m_{M} \vert \psi_{ \mathrm{out}} \rangle \nonumber \\
         =& \frac{1}{\prod^{M}_{i=1} (m_{i}!)}\nonumber\\
         &~\frac{\partial F(a^{\dagger}_{1}, \dots , a^{\dagger}_{N})}{\partial a^{\dagger m_{1}}_{(N-M+1)} a^{\dagger m_{2}}_{(N-M+2)}\dots a^{\dagger m_{M}}_{N}} \Bigg|_{\substack{a^{\dagger}_{(N-M+1)}=0 \\ \dots \\ a^{\dagger}_{(N)}=0}} \vert 0 \rangle, 
\end{align}  
where $G(a^{\dagger}_{1}, \dots , a^{\dagger}_{N-M})$ is a homogeneous polynomial of degree 
$n-m$ in the creation operators $a^{\dagger}_{i}$, here $1 \leq i \leq (N-M)$. The coefficients of the monomials in $G$ are again homogeneous polynomials of degree $n$ belonging to the polynomial ring $\mathbb{C}[A_{i,j}]$.

We can also represent the target state vector as a homogeneous polynomial
\begin{align}
        \vert \psi_{ \mathrm{tar}} \rangle & := Q(a^{\dagger}_{1}, \dots , a^{\dagger}_{N-M}) \vert 0 \rangle.  
\end{align} 
The equivalence of the heralded state from the circuit and the target state can hence be reduced to the equivalence of the corresponding polynomials.  Solving $\vert \psi_{ \mathrm{post}} \rangle = \vert \psi_{ \mathrm{tar}} \rangle$ would be equivalent to solving $G=Q$. Note, however, that $\vert \psi_{ \mathrm{post}} \rangle$ is an unnormalized state vector. The correct normalization is determined by the probability of measuring the heralding pattern $(m_{1},\dots,m_{M})$. 
Therefore, the equivalence means solving the equations 
\begin{equation}
    \vert \psi_{ \mathrm{post}} \rangle = \alpha \vert \psi_{ \mathrm{tar}} \rangle \Longleftrightarrow	 G = \alpha Q, 
\label{eq:equivalence}
\end{equation}
where $\alpha \in \mathbb{C}, \alpha \ne 0$ and $\vert \alpha \vert^{2}$ determines the heralding success probability. To directly ensure the condition of a nonzero success probability, $\alpha\ne0$, in our work we rewrite this as \begin{equation} \label{eq:main_stategeneqn}
    \gamma \vert \psi_{ \mathrm{post}} \rangle =  \vert \psi_{ \mathrm{tar}} \rangle \Longleftrightarrow	\gamma G =  Q, 
\end{equation} 
where $\gamma=1/{\alpha}\in\mathbb{C}$.

From $\gamma G =  Q$, one can equate the corresponding monomial coefficients to derive a system of polynomial equations where the variables are the elements of the unknown transformation matrix $A$. Although this system can be solved using the Gr\"{o}bner basis technique, doing so is computationally expensive, with a worst-case time complexity that is doubly exponential 
\cite{rolnick2019robust}. Moreover, the Gr\"{o}bner basis technique provides more information than just whether a solution exists. Therefore, when the focus is on feasibility, we propose instead using the methods described in the following Section \ref{infeasibile} to determine whether $\gamma G = Q$ has a solution or not.
 
\subsection{Computing infeasibility certificates for heralded photonic state generation} \label{infeasibile}

The NulLA algorithm developed in Ref.~\cite{DELOERA20111260} to prove the infeasibility of a system of polynomial equations, $f_1(x) = 0$, ~\dots, $f_s(x) = 0$,
is based on the \emph{Weak Nullstellensatz} 
\cite{AlgebraicGeometry, TaoBlog} as a reading of Hilbert's Nullstellensatz. We use a simplified version, adapted from Ref.~\cite{DELOERA20111260} as relevant to our situation of algebraically closed fields:

\begin{theorem}[Variant of Hilbert's Nullstellensatz]
Let $\mathbb{K}$ be an algebraically closed field. Given $f_1, f_2,\dots, f_s \in \mathbb{K}[x_1,\dots, x_n]$, the system of polynomial equations $f_1(x) = 0$, ~\dots, $f_s(x) = 0$, has no solution in ${\mathbb{K}}^n$ if and only if there exist polynomials $\beta_1,\dots, \beta_s \in \mathbb{K}[x_1,\dots, x_n]$ such that $1 =\sum_{i=1}^s \beta_i(x)f_i(x)$. 
\end{theorem}

The polynomial identity $1 = \sum_i \beta_i f_i$ is called a Nullstellensatz certificate, which has degree $d$ if $\mathrm{max}_i\{\mathrm{deg}(\beta_i)\} = d$.

One can attempt to find a Nullstellensatz certificate of a specific degree $d$ by setting up and solving a system of linear equations. This system is constructed as follows: For each $f_i$, one sets up a general polynomial $\beta_i$ of degree $d$ belonging to the polynomial ring $\mathbb{C}[x]$, with the coefficients of $\beta_i$ represented by unknowns $c$. Since two polynomials are identical if and only if their corresponding monomial coefficients are equal, the equation $1=\sum_i\beta_i f_i$ can be grouped by monomials in $x$, and each monomial yields a (linear) equation for the unknowns $c$. If a solution to the linear system exists, it provides an infeasibility certificate.

In principle, the NulLA algorithm can determine whether a solution to the original system of polynomial equations exists or not. It does so by attempting to find a Nullstellensatz infeasibility certificate of minimal degree $d=\mathrm{max}_i\{\mathrm{deg}(\beta_i)\}$, starting at $d=1$. If unsuccessful, it increments the degree until either a certificate is found or an upper bound on the necessary degree is reached. If such a certificate is found, the original system is infeasible. Otherwise, the system is deemed feasible.
In practice, however, reaching the upper bound on the possible necessary degree, which would guarantee a decision on the feasibility problem, is not generally computationally attainable (see Section~\ref{sec:properties}). Regardless, it can be possible to find infeasibility certificates with much lower degrees than the upper bounds ~\cite{DELOERA20111260}. We therefore propose using the NulLA algorithm to establish no-go results and lower bounds on resource requirements within the context of heralded state generation.

The problem of checking if a given target state can be reached from a given input state and heralding pattern can be reduced to checking for the existence of a Nullstellensatz infeasibility
certificate for the system of polynomial equations obtained from Eq.~(\ref{eq:main_stategeneqn}).
The NulLA algorithm employs a distinct approach compared to the Gr\"{o}bner basis techniques. 
The Gr\"{o}bner basis techniques simultaneously determine whether a solution exists and what that solution is. By contrast, NulLA focuses on the sole question of whether a solution exists and functions progressively, by incrementally checking for infeasibility certificates of increasing degrees.

\section{Properties of the technique \label{sec:properties}}
This section introduces key simplifications for establishing the infeasibility of state generation protocols, given a fixed set of resources. Specifically, when starting with a fixed number of available input photons, we show that it is sufficient to test just a single configuration of input and heralding photons, rather than checking all possible configurations. Afterwards, we present an analysis of the scaling behavior of NulLA for heralded state generation.

\subsection{Problem simplification}
Here, we focus on Fock state inputs and heralding patterns with total photon numbers $n$ and $m$, respectively. We discuss the use of input states without a well-defined photon number in Section \ref{subsec:probsources}. Our framework considers a linear transformation over $N$ modes, of which $M$ modes are designated for heralding. In our analysis, we focus on auxiliary photons as the resource of interest---in contrast to vacuum input modes and vacuum heralding, which may, for example, be added later on to render a transformation unitary. 
In Lemma \ref{lem:optimal inputs} we identify the optimal input states and heralding patterns, where `optimal' is defined as the most powerful for proving infeasibility, in the following sense: If the state generation problem is infeasible for the optimal configuration, then it is also infeasible for any other configuration of Fock state inputs and heralding patterns with the same total number of photons.
This lemma simplifies the state generation problem by enabling a categorization based solely on the number of available input photons $n$ and target state vector $\vert \psi_{ \mathrm{tar}} \rangle $, noting that these two determine the number of heralding photons $m$. That is, for a given target state and choice of $n$, only one configuration of input state and heralding pattern needs to be 
tested to prove infeasibility.

\begin{figure}[htbp!]
\centering
\subfloat[\label{fig:inp_actual}]{\includegraphics[width=0.9\columnwidth]{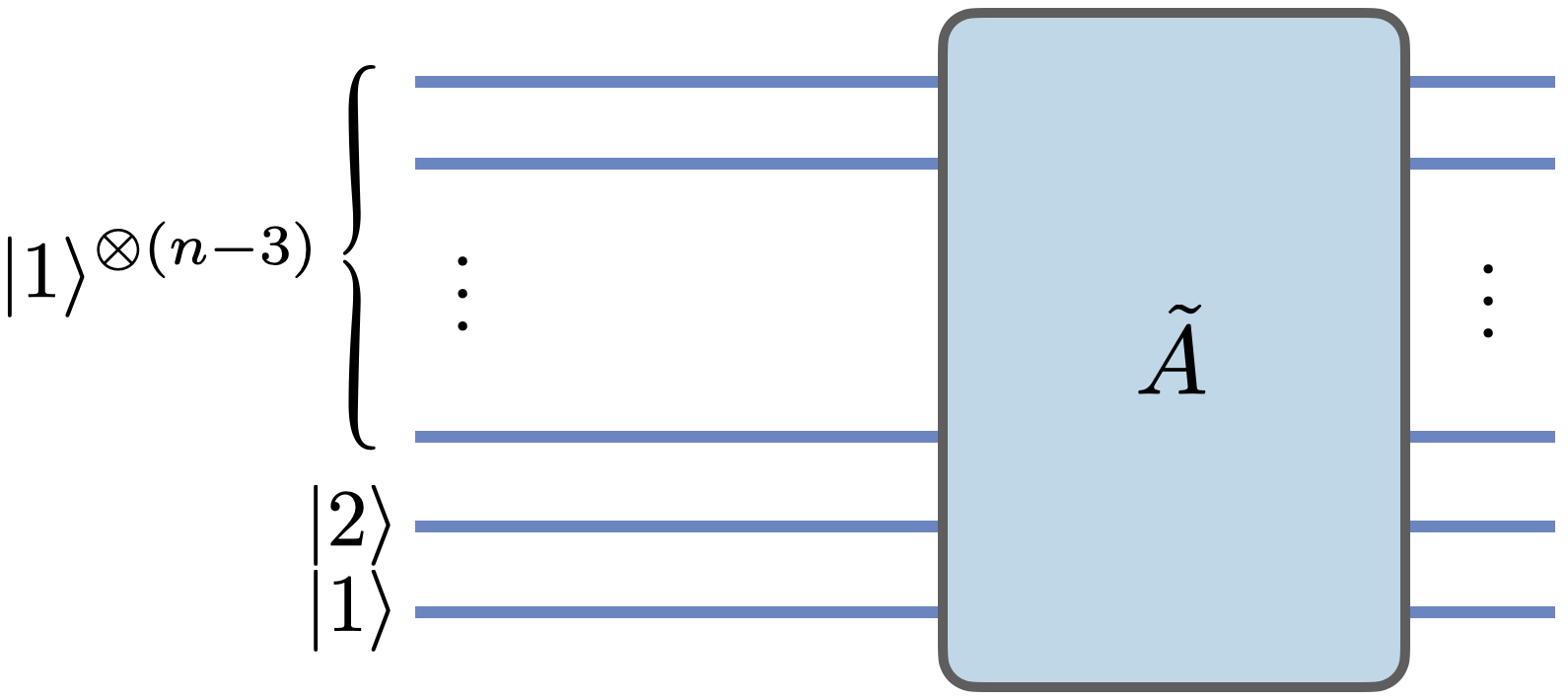}}\vfill
\subfloat[\label{fig:inp_optimal}] {\includegraphics[width=0.94\columnwidth]{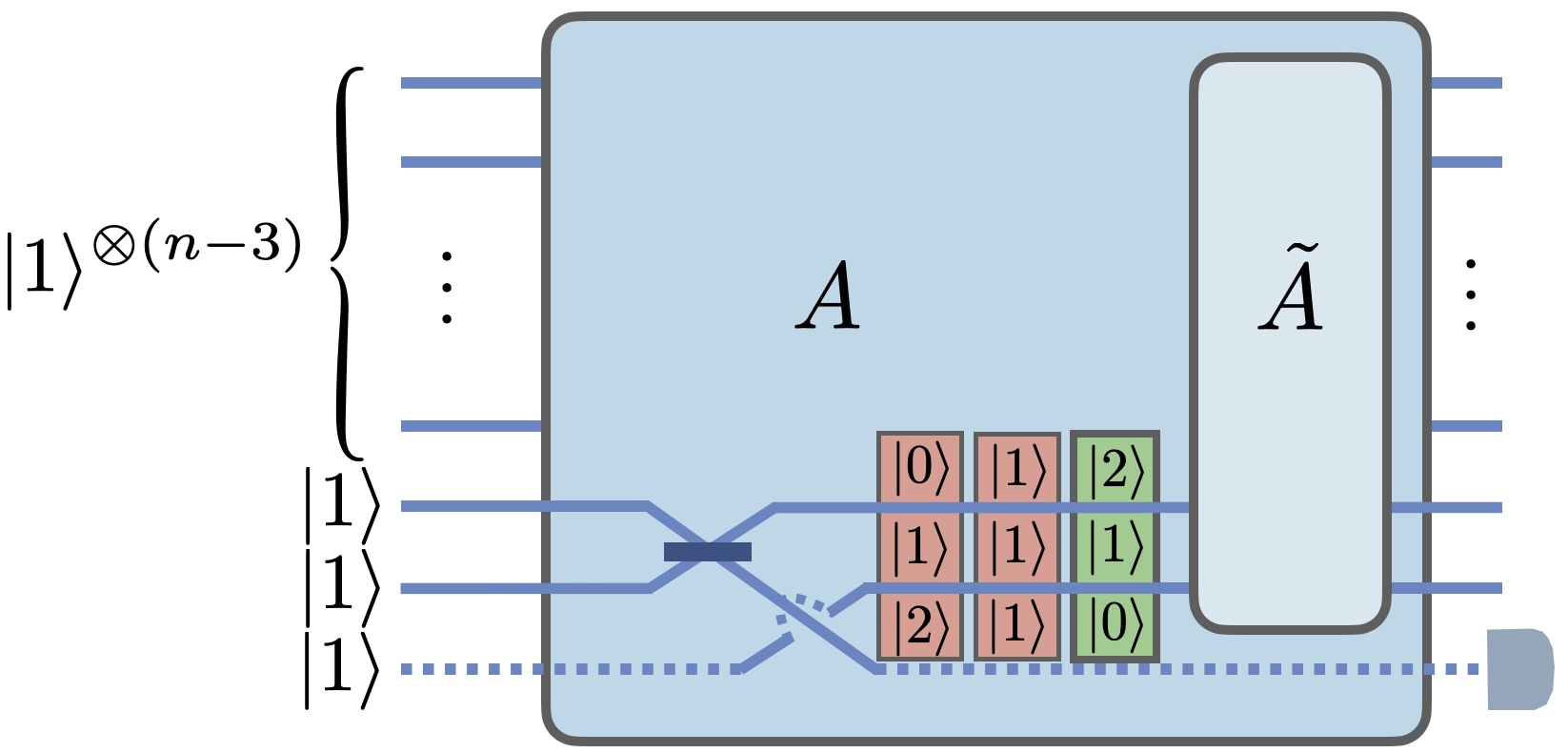}}
\caption{Example demonstrating the optimality of single-photon inputs as described in Lemma \ref{lem:optimal inputs}. The scenario depicted in Fig.~\ref{fig:inp_actual} involves state generation tasks requiring $n-3$ single photons in the top $n-3$ modes of an interferometer $\tilde{A}$, with two photons and one photon in the remaining two modes, respectively. We demonstrate that this setup can be replaced by the configuration shown in Fig.~\ref{fig:inp_optimal}, where a larger interferometer $A$ achieves the generation of the same state using only single-photon inputs and heralding on vacuum in the newly introduced ancilla mode. The green pattern represents 
the desired configuration, while the red patterns indicate failed configurations that are ruled out by heralding.}

\label{fig:inp_heralding}
\end{figure}

\begin{lemma}[Optimal input states and heralding patterns]
Given $n$ input photons and $m$ heralding photons, it is \emph{sufficient} to consider only the input state vector $\left|\psi_{\mathrm{in}}\right\rangle=\prod^{n}_{i=1}a^{\dagger}_{i} \left|0\right\rangle$ and the heralding pattern $(m_{1},m_{2}\dots,m_{m})$=$(1,1,\dots,1)$ to prove infeasibility of all separable $n$-photon input states and all heralding patterns containing $m$ photons. 
Therefore, we consider these the \emph{optimal} input and heralding patterns. 
\label{lem:optimal inputs}
\end{lemma}

To see that the the input state vector $\prod^{n}_{i=1}a^{\dagger}_{i} \left|0\right\rangle$ is optimal, consider the ability to add a photon to mode $i$ from a single-photon input in mode $j$, by using a beam splitter between modes $i$ and $j$, and subsequent vacuum heralding on mode $j$.
An arbitrary, pure, $n$-photon Fock state input 
\begin{equation}
    |\tilde{\psi}_{\mathrm{in}}\rangle=\prod^{N}_{i=1}\frac{1}{\sqrt{\tilde{n}_i!}} a^{\dagger \tilde{n}_i}_{i}\left|0\right\rangle
\end{equation}
can be built out of the state vector $\prod^{n}_{i=1}a^{\dagger}_{i} \left|0\right\rangle$ by repeating this photon-addition action $(\tilde{n}_i-1)$ times with single-photon input modes, to build up each desired input mode with associated $\tilde{n}_i> 1$. As a 
simple example, the process for generating the state vector $\left|2,1\right\rangle$ can be seen within Figure \ref{fig:inp_heralding}. All the vacuum heralding events from these photon additions can be postponed until the end of the circuit and included in the heralding pattern, and the beam splitters can be absorbed into the unknown transformation $A$. In this way, the optimal input state vector $\left|\psi_{\mathrm{in}}\right\rangle=\prod^{n}_{i=1}a^{\dagger}_{i} \left|0\right\rangle$ covers all other Fock-state inputs $|\tilde{\psi}_{\mathrm{in}}\rangle$. 

\begin{figure}[htbp!]
\centering
\subfloat[\label{fig:out_actual}]{\includegraphics[width=0.93\columnwidth]{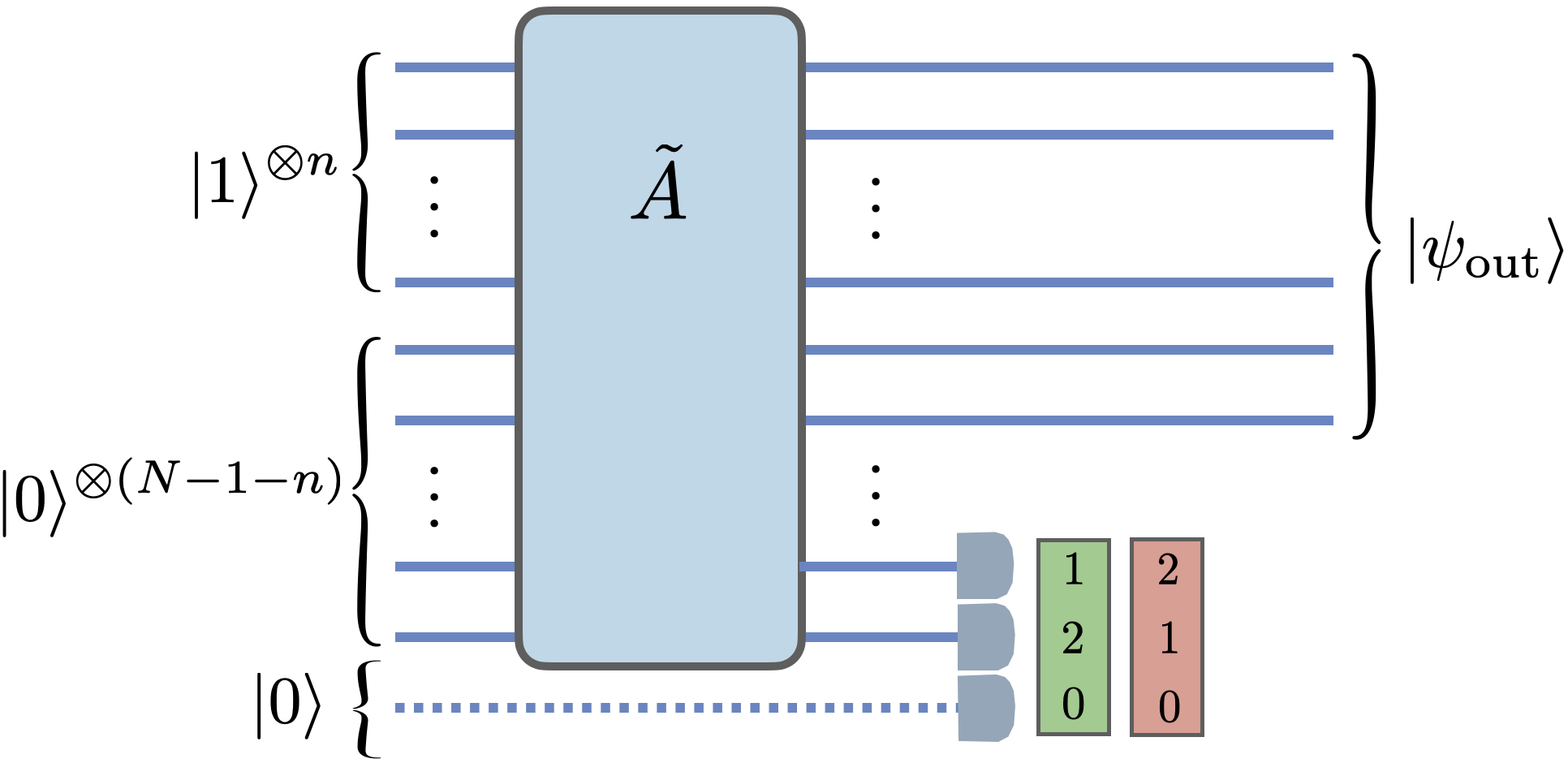}}\vfill
\subfloat[\label{fig:out_optimal}]{\includegraphics[width=0.91\columnwidth]{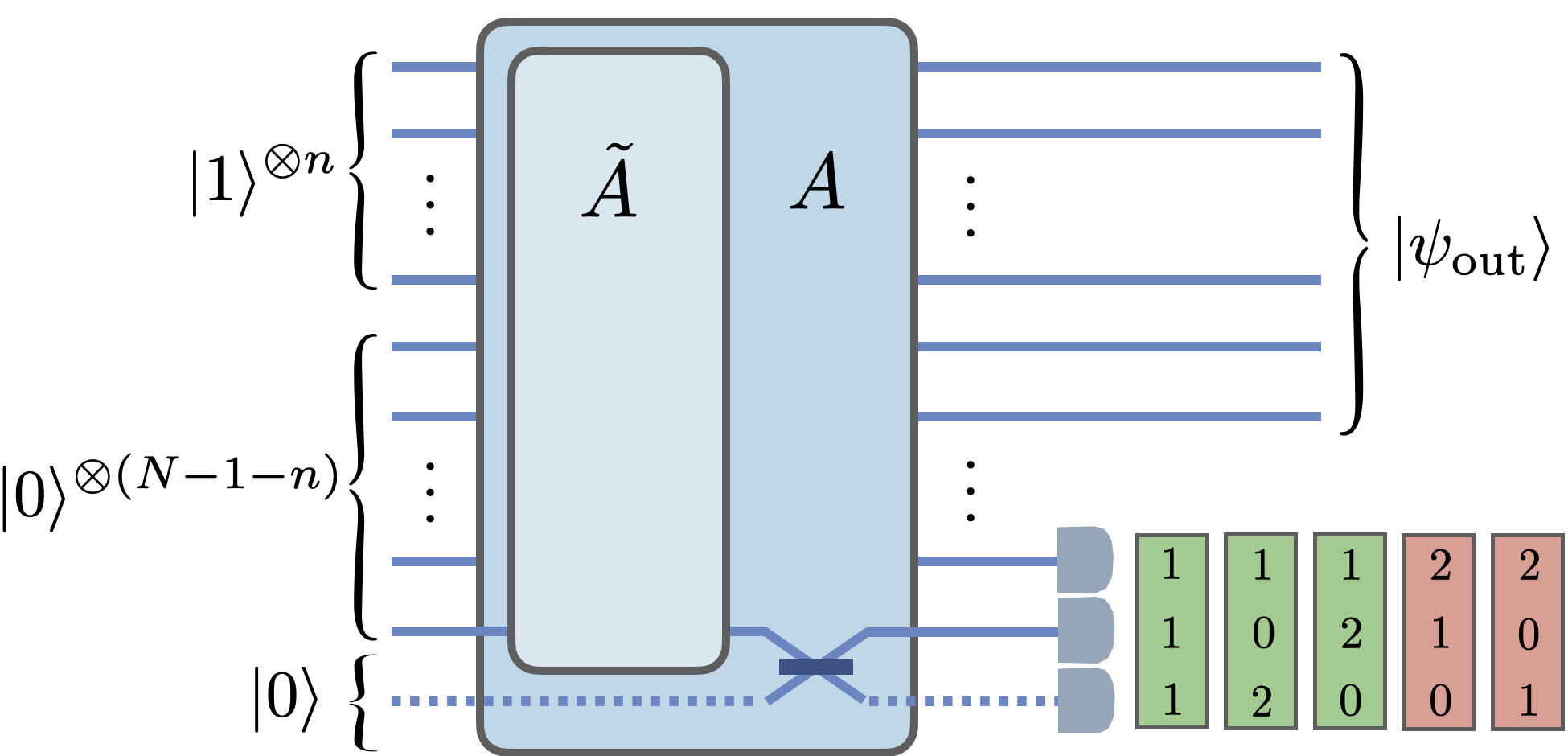}}\vfill
\caption{Example demonstrating the optimality of the single-photon heralding pattern as described in Lemma \ref{lem:optimal inputs}. The scenario shown in Fig.~\ref{fig:out_actual} involves state generation tasks requiring an input of $n$ single photons in the top $n$ modes of an interferometer $\tilde{A}$, with vacuum in $N-1-n$ modes. The dashed bottom mode below $\tilde{A}$ is there for illustration purposes when comparing with Fig.~\ref{fig:out_optimal} only; it is superfluous for the state generation and can be disregarded. The desired outcome occurs when three photons occupy the last two modes, following a specific heralding pattern where one photon is in the second-to-last mode and two photons are in the last mode. This setup can be replaced by the one in Fig.~\ref{fig:out_optimal}, where a larger interferometer $A$ achieves the generation of the same state using a single-photon heralding pattern. The green patterns indicate the possible photon-number measurement outcomes at the detectors that are associated with the successful heralding pattern of Fig.~\ref{fig:out_actual}, with the first one being the optimal single-photon heralding pattern. The red patterns indicate the possible photon-number measurement outcomes at the detectors that are associated with the wrong heralding pattern of Fig.~\ref{fig:out_actual}; these are successfully rejected by the single-photon heralding pattern.
}
\label{fig:out_heralding}
\end{figure}

To understand why the optimal heralding pattern is $(1,1,\dots,1)$, a result previously discussed in Ref.~\cite[p.~51]{kok2001statepreparationquantumoptics}, assume we have an original transformation $\tilde{A}$ and some arbitrary original heralding pattern $(\tilde{m}_1,\tilde{m}_2,\dots, \tilde{m}_M)$ that work. Now the idea is to append a fan-out to the transformation and use the heralding pattern $(1,1,\dots,1)$ afterwards. We illustrate this strategy in Fig.~\ref{fig:out_heralding}, which shows how the heralding pattern $(1,2)$ can be replaced by introducing an ancilla mode and applying the heralding pattern $(1,1,1)$ instead.

Given that $(\tilde{m}_1,\tilde{m}_2,\dots, \tilde{m}_M)$ is the photon distribution that would correctly herald the target state after $\tilde{A}$, we need to be able to transform it into something that has nonzero probability of being detected by $(1,1,\dots,1)$, while also ensuring that all the other, incorrect photon distributions after $\tilde{A}$ have zero probability of being detected by $(1,1,\dots,1)$ at the end. 
If $(\tilde{m}_1,\tilde{m}_2,\dots, \tilde{m}_M)=(1,1,\dots,1)$, this is already the case and there is nothing to do. For any other chosen pattern $(\tilde{m}_1,\tilde{m}_2,\dots, \tilde{m}_M)$, this can always be achieved with a (or a sequence of) beam splitter(s) serving as a fanning out operation. If there was a nonzero probability of getting the original heralding pattern $(\tilde{m}_1,\tilde{m}_2,\dots, \tilde{m}_M)$ after $\tilde{A}$, then one can choose a circuit such that there is a nonzero probability of getting $(1,1,\dots,1)$ at the end. This circuit is $\tilde{A}$ followed by a sequence of beam splitters. These beam splitters combine the multi-photon modes of the original heralding pattern with vacuum modes to probabilistically distribute the photons into multiple modes.
For the same circuit, any incorrect photon distributions after $\tilde{A}$ should have zero probability of being detected by the heralding pattern $(1,1,\dots,1)$ at the end. This is true because any incorrect distribution must have fewer photons than $(\tilde{m}_1,\tilde{m}_2,\dots, \tilde{m}_M)$ in at least one of the modes, and this makes it impossible to obtain a photon in every mode after fanning out.

This strategy of constructing $A$ works for any chosen pattern after $\tilde{A}$, although the specific appended transformation depends on the choice of $(\tilde{m}_1,\tilde{m}_2,\dots, \tilde{m}_M)$. Thus, for any chosen heralding pattern $(\tilde{m}_1,\tilde{m}_2,\dots, \tilde{m}_M)$ after the transformation $\tilde{A}$, there exists another transformation $A$ for which the heralding pattern $(1,1,\dots,1)$ can be used instead.

Note that only the existence of a solution is considered here. There is no guarantee that the configurations from Lemma \ref{lem:optimal inputs} provide the maximal possible success probability. However, when focusing on the question of feasibility, this approach allows proving infeasibility for a given number of input photons after testing a single configuration. 
It effectively reduces the problem to two independent variables: the target state and the number of heralding photons. The total number of modes $N$ to be included in the transformation $A$ that is used to prove infeasibility of the heralded quantum state generation task is
\begin{equation}
 N=\max\left\{n,N_{\mathrm{T}}+m\right\},
\end{equation}
where $N_{\mathrm{T}}$ is the number of modes in the target state. Putting this all together, we can deduce
the following.

\begin{theorem}[Infeasibility theorem]
The infeasibility of generating any pure $(n-m)$-photon, $N_T$-mode target state from a multi-mode Fock state by heralding $m$ photons 
can be proven by utilising the NulLA algorithm, considering only the input state and heralding pattern as specified in Lemma \ref{lem:optimal inputs}. 
\end{theorem}

\subsection{Scaling}\label{subsec:scaling}
In this section, we provide upper bounds for the size of the linear system obtained in the NulLA algorithm as a function of the chosen test degree $d$, as well as an upper
bound on the maximal degree of infeasibility certificates one might need to test to conclusively decide feasibility---that is, a bound on the degree that might be required to prove infeasibility, and which is thereby sufficient for proving feasibility. Appendix \ref{app:scaling} provides more details on how the following scaling results were obtained. 

\emph{Number of polynomial equations}: The number of polynomial equations is ${n-m+N-M-1 \choose n-m}$.

\emph{Number of unknowns in the linear system}: Let $V_{\mathrm{max}}:=\mathrm{min}(Nn,N^2)$. 
The number of unknowns in the linear system of the NulLA algorithm is upper bounded by ${n-m+N-M-1 \choose n-m}{V_{\mathrm{max}} + d \choose d}$. 

\emph{Number of equations in the linear system}:
\begin{itemize}
\item If $d\ge n-1$, 
the number of equations in the linear NulLA system is upper bounded by $V_{\mathrm{max}}+d+n \choose V_{\mathrm{max}}$.
\item If $d< n-1$, 
the number of equations in the linear system is upper bounded by ${V_{\mathrm{max}}+d+n \choose V_{\mathrm{max}}}-{V_{\mathrm{max}}-1+n \choose V_{\mathrm{max}}}+{V_{\mathrm{max}}+d \choose V_{\mathrm{max}}}$.
\end{itemize}

\emph{Upper bounds on sufficient infeasibility degree}: Upper bounds on the infeasibility degree that is sufficient to decide feasibility are given by Koll\'{a}r and Sombra \cite{Kollar88,Sombra99}. Let there be a system of $s$ polynomial equations, $F=\{f_1=0,\dots,f_s=0\}$, in descending order of polynomial degrees such that $d_1\ge...\ge d_s$, with $V\ge 2$ variables. An upper bound on the sufficient infeasibility certificate degree, $K$, which is optimal if $d_s>2$, is provided by Koll\'{a}r, whereas Sombra's bound may provide an improvement outside this regime. Koll\'{a}r's bound, which we focus on here for simplicity, and where we restrict ourselves to the case $d_s\ge3$, 
is
\begin{equation}
K\le d_s\Pi _{j=1}^{\mathrm{min}(V,s)-1}d_j-d_s.
\end{equation}
To apply this bound in the context of state generation, we can use that 
$V\le V_{\textrm{max}}= \mathrm{min}(Nn,N^2)$, 
while $s={n-m+N-M-1 \choose n-m}$ 
and $d_i=n$ $\forall i$ to get
\begin{equation}
K\le n^{\mathrm{min}\{V_{\textrm{max}},s\}}-n.
\end{equation}

While this upper bound scales poorly in the system size [for example, it is $O(10^{17})$ when $n=4$, $N=4$ and $m=1$], in practice, infeasibility certificates are typically found well below the 
bound, as we demonstrate by example applications in the next section. One can try to further simplify the computation through the use of equivalence classes. However, as discussed in Appendix \ref{app:equivalenceclasses}, this can sometimes yield misleading results and so is not considered in this text.

\section{Applications \label{sec:applications}}

In this section, we apply the NulLA algorithm to examples of state-generation tasks to explore typical behavior. We demonstrate how the NulLA algorithm can be useful for proving the resource optimality of known heralding schemes and automated search processes for families of target states. Furthermore, we discuss how our approach can be extended to address the design of gates and the use of probabilistic sources. 

\subsection{Resource bounds for Bell state generation}\label{sec:Bell}
Solutions for dual-rail encoded Bell state generation from multi-mode Fock states are known for four-photon inputs, e.g., Refs.~\cite{Zhang2008, Carolan2015, Gubarev2020}. Applying the NulLA technique to the three-photon input $\vert 1,1,1,0,0 \rangle$ with the heralding set to $(1)$ and the target state set to the Bell state vector $(\vert 1,0,1,0 \rangle+\vert 0,1,0,1 \rangle)/{\sqrt{2}}$ provides an infeasibility certificate with degree 9. 
This proves that generating a Bell state from a separable three-photon input state is impossible, thereby confirming that the known solutions utilizing four photons are optimal in terms of the number of required input photons. The above proof via the NulLA algorithm is consistent with a result in Ref.~\cite{Stasja2017}, where the specific case of Bell state generation was analyzed on an individual basis. However, the systematic nature of our approach allows for an automated process that can be used to easily tackle a large number of problems, as we show below.

\subsection{Infeasibility certificate degrees for wider families of two-photon target states}
The above test for the Bell state reveals an infeasibility certificate already at degree 9, despite the upper bound being 59046. This vast difference between the worst-case scenario and the degree required in practice raises the question of whether this is an isolated example or a more generic feature. The Bell state possesses particular symmetries, and these could plausibly give rise to unusual features. Hence, we test the behavior of NulLA on hundreds of random two-photon target states that do not share the Bell state's symmetries, to explore if the desirable feature of a large gap persists. 

We apply the algorithm to state generation tasks that can be divided into four groups. The four groups, presented as rows in Table \ref{tab:haar random states IC}, are based on the number of modes in the target state and the number of heralding photons. We explore state generation tasks with zero and one heralding photons. For each of these choices, we use two sets of 200 random target states to explore two-photon target states involving three modes and four modes, respectively. To create the input states, the input photons are distributed according to the simplification from Lemma \ref{lem:optimal inputs}. Consequently, the inputs for these four groups are: $\vert 1,1,0 \rangle$ for preparing two-photon target states in three modes with no heralding photon, $\vert 1,1,0,0 \rangle$ for two-photon target states in four modes with no heralding photon, $\vert 1,1,1,0 \rangle$ for two-photon target states in three modes with one heralding photon, and $\vert 1,1,1,0,0 \rangle$ for two-photon target states in four modes with one heralding photon. 

Using multi-mode Fock states as basis states, two-photon states across three and four modes can be described as linear combinations of six and ten basis states, respectively. The Bell state is a specific example of a two-photon state across four modes, where all but two of the ten Fock states have zero contribution, and the nonzero contributions of the remaining two Fock states are equal. 
To obtain more generic states as the target states, we create 200 Haar-random two-photon states across three modes and another 200 Haar-random two-photon states across four modes. 
Each set of 200 target states is used for two of the four groups, once to form the tasks with zero heralding photons, and once more for the tasks with one heralding photon.

The results are presented in Table \ref{tab:haar random states IC}. Of the four groups, one represents a state generation task that is known to always be feasible, namely the generation of target states that consist of two photons in three modes when using one heralding photon~\cite{degliniasty2024SimpleRulesTwoPhoton}. 
For this set of state generation tasks, no infeasibility certificates are found in tests up to degree nine; this is consistent with the feasibility of the task but does not prove it, since one would need to test up to degree 726, which is computationally out of reach. For each of the three remaining groups, infeasibility certificates are found for the randomly generated target states, thereby proving that the corresponding state generation tasks are impossible. These infeasibility certificates are always found at a fixed value of the certificate degree in a given group. 
Again, we observe infeasibility certificates at low degrees compared to the upper bound $K$. This demonstrates that for Haar-random states, extremely large gaps persist between the worst-case behaviour and the degree that suffices in practice.

\begin{table}
    \centering
    \setlength{\tabcolsep}{1mm}
    \begin{NiceTabular}{c@{\hspace{5mm}}ccccc}[colortbl-like]
        \hline
        \RowStyle{\bfseries}
        \thead{Target\\modes,\\$N_{\mathrm{T}}$} & \thead{Heralding\\photons,\\$m$} & \thead{Input\\photons,\\$n$} & \thead{Total\\modes,\\$N$} & \thead{Computed\\degree,\\$K^*$} & \thead{Upper\\bound on\\$K$} \\
        \hline\hline
        3 & 0 & 2 & 3 & 4 & 126 \\
        \rowcolor{gray!20}
        3 & 1 & 3 & 4 & - & 726  \\ 
        4 & 0 & 2 & 4 & 4 & 510 \\ 
        \rowcolor{gray!20}
        4 & 1 & 3 & 5 & 9 & 59046 \\
        \hline
    \end{NiceTabular}
    \caption{
    Infeasibility certificate results for generating two-photon Haar-random states in three or four modes, using either zero or one heralding photon.
    Within each group, we found that the degree of the computed infeasibility certificate, $K^*$, remained constant across all target states. The group with no infeasibility certificates returned (up to a degree 9) is known to correspond to feasible state-generation tasks. The upper bounds on the sufficient infeasibility certificate degree, $K$, are derived from Refs.~\cite{Kollar88, Sombra99}.
    }
    \label{tab:haar random states IC}
\end{table}

\subsection{Inputs for vacuum-heralded NOON state generation}

N00N states are entangled two-mode states with a definite photon number, defined as 
\begin{equation}
     \ket{\mathrm{N00N}}_n = \frac{1}{\sqrt{2}} (\ket{n,0} + \ket{0,n}),
     \label{eq:def_noon_state}
 \end{equation}
 where we adopt the lower-case $n$ rather than the more conventional upper-case $N$ as the total photon number to align with our existing definition of the number of input photons, for reasons that will become evident soon.
 These states are well-known for being the optimal probes in phase sensing tasks without photon loss \cite{PhysRevA.54.R4649,PhysRevLett.102.040403,demkowicz-dobrzanski_elusive_2012}.

 An interesting property of NOON states is the possibility of generating them with vacuum heralding alone, that is, with zero heralding photons $(m=0)$. Indeed, several vacuum-heralded schemes are known for generating NOON states, all of which use an input of $n$ single photons separated into $n$ modes \cite{PhysRevA.65.053818,LEE2012307,PhysRevA.68.052315,PhysRevA.70.023812,forbes2025heraldedgenerationentanglementphotons}. Based on Lemma \ref{lem:optimal inputs}, this is the most powerful input, and it may or may not be possible to redistribute the input photons. It is, therefore, natural to ask if the input photons indeed \emph{need} to be spread out to the single-photon level, or if a vacuum-heralded scheme can also exist that has at least two input photons in the same mode.

 To address this question, we analyze the state generation task of creating the target state vector $\vert\psi_{ \mathrm{tar}} \rangle=(\ket{3,0} + \ket{0,3})/\sqrt{2}$ from the input state vector $\vert \psi_{ \mathrm{in}}\rangle=\vert2,1\rangle$, and similarly for larger $n$, the state generation tasks of creating $\vert\psi_{ \mathrm{tar}} \rangle= (\ket{n,0} + \ket{0,n})/\sqrt{2}$ from the $(n-1)$-mode input state vector 
 \begin{equation}
 \vert \psi_{ \mathrm{in}}\rangle=\vert2,1,\dots,1 \rangle.
 \end{equation}
 Explicit vacuum heralding is included on $n-3$ modes. Further vacuum heralding modes could arise when a non-unitary transformation is rendered unitary by including additional modes, but those modes do not need to be accounted for here. Testing NOON states with $n=\left(3,4,5,6,7\right)$, we find infeasibility certificates with degree 3,4,5,6, and 8, respectively.
 This result establishes the need to spread out the input photons to the single-photon level, and illustrates that these inputs are the most powerful, as per Lemma \ref{lem:optimal inputs}.

\subsection{Transformations for sets of input--output state pairs}

The focus of this work has been on the feasibility of heralded state generation, that is, a transformation between one input state and one output state. However, the same approach can be broadened to the feasibility of other heralded transformations, where a single transformation relates not just one but a \emph{set} of input--output state pairs. Such transformations between sets of input--output state pairs can be used to model different objects of interest in optical quantum information processing, including heralded quantum gates and quantum non-demolition measurements. 

The previously described method only requires a minor adaptation: Instead of a single polynomial equation in creation operators as in Eq.\ \eqref{eq:main_stategeneqn}, we now obtain a system of such polynomial equations 
\begin{eqnarray}
    \label{eq:multiple_stategeneqn}
    \gamma G_1 &=&  Q_1 ,\\ &\vdots& \nonumber \\ \gamma G_k &=&  Q_k \nonumber,
\end{eqnarray} 
where each equation stems from a pair of input--output states. Similarly as in the state generation task, one can equate the corresponding monomial coefficients for each polynomial equation in creation operators and, because the same unknown transformation has to simultaneously satisfy 
all the input--output relations, this 
creates one large system of polynomial equations where the variables are the elements of the unknown transformation matrix $A$. The feasibility of this system of polynomial equations can now be tested via the NulLA algorithm as before.

\subsubsection{Example: Minimal resource requirements for the heralded CNOT gate}

We demonstrate the above approach on the CNOT gate, a cornerstone of quantum information processing. When used on dual-rail-encoded photonic qubits---and in the absence of additional heralding photons---a CNOT gate is a two-photon, four-mode transformation that maps the logical basis as $\vert1,0,1,0\rangle \mapsto \vert1,0,1,0\rangle$, $\vert1,0,0,1\rangle \mapsto \vert1,0,0,1\rangle$, $\vert0,1,1,0\rangle \mapsto \vert0,1,0,1\rangle$, and $\vert0,1,0,1\rangle \mapsto \vert0,1,1,0\rangle$. There are known schemes for heralded CNOT gates that use two heralding photons. For instance, Ref.~\cite{Carolan2015} implements a four-photon, six-mode transformation, which can be cast as four simultaneous state generation tasks: i.e., obtaining the target state vector $\vert\psi_{\mathrm{tar}} \rangle=|1,0,1,0\rangle$ from the input state $\vert \psi_{\mathrm{in}} \rangle=|1,0,1,0,1,1\rangle$, and similarly for the rest of the basis states mentioned above. Measuring one photon in each of the last two heralding modes, corresponding to the heralding pattern (1,1), flags the success of the probabilistic gate. 

Our approach allows testing the resource-optimality of this known solution. Using the NulLA algorithm, we successfully obtain an infeasibility certificate with degree $d=6$ for the case where only a single heralding photon is used. Specifically, we test the three-photon, five-mode transformation derived from the original scheme when one input photon is removed, and where the heralding pattern reduces to (1). This result ascertains that at least two ancilla photons are necessary to herald a CNOT gate.

\subsection{Extension to probabilistic photon sources}\label{subsec:probsources}

The analysis and examples in this article have taken the input states to be product states with a fixed number of input photons, as would naturally be obtained from deterministic photon sources. However, probabilistic sources, e.g., probabilistic photon-pair sources based on non-linear processes such as spontaneous parametric down-conversion, are another popular choice for generating the input photons for heralded state generation tasks. Such probabilistic sources may produce different numbers of photons (e.g., even numbers in the case of probabilistic photon-pair sources). Moreover, if several probabilistic sources are coherently pumped, then this usually also produces a superposition of different inputs, even when fixing the number of photons. 

Our approach can be adapted to account for probabilistic sources as follows.
Different photon numbers produced by the source can be treated one by one and subsequently put together as a set of state-generation tasks that have to be achieved with a single transformation. For a given photon number, $\vert \psi_{ \mathrm{in}} \rangle$ may not be a product state vector if several probabilistic sources are coherently pumped, and $\vert \psi_{ \mathrm{out}} \rangle$ needs to be calculated accordingly. For the case of the photon number that corresponds to the total number of photons expected after the transformation (the sum of target photons and heralding photons), the probability of obtaining a heralding signal needs to be nonzero, and the heralded state needs to be the target state, as usual. By contrast, for the case of any photon number that does not correspond to the total number of photons expected after the transformation, the probability of obtaining a heralding signal needs to be zero. This is because the heralded state would have a different number of photons than the target state, thereby reducing the fidelity of the heralded state with the target state. Since probabilistic sources emit higher numbers of photons with decreasing probabilities, one would typically choose to include numbers of photons up to and including the total photons expected after the transformation, and consider higher-order terms as an error that can be reduced in practice by reducing the pump power, at the expense of lower count rates.

\section{Conclusion \label{sec:conclusion}}

Sophisticated photonic quantum technologies require the preparation of complex quantum states. For example, it is known that the availability of GHZ states lessens the burden of conditioned multiplexed quantum dynamics for linear-optical quantum computing \cite{PhysRevLett.99.130501}. In this work, we have reconsidered the problem of preparing photonic quantum states from the perspective of algebraic geometry.
We have shown that the NulLA algorithm provides a rigorous and powerful tool for proving the infeasibility of generating a target state using linear optics augmented with heralding. When an infeasibility certificate is found, this conclusively \emph{proves} infeasibility, unlike when standard gradient-based numerical optimization techniques fail to find a solution. Moreover, our approach provides a targeted test of infeasibility of a state generation task, unlike numerical optimization techniques and Gr\"{o}bner basis techniques that inherently go beyond the decision problem and attempt to provide additional information about the solutions. 

We demonstrate our approach on a number of different examples of state generation tasks. While the degree of the infeasibility certificate attempt that is required to prove feasibility is, in general, prohibitive, we show that to prove \emph{infeasibility}, it often suffices to attempt infeasibility certificates of low degrees. This observation
is consistent with findings of the NulLA algorithm applied in other contexts \cite{DELOERA20111260}. 
We have further discussed how probabilistic input sources can be included, and shown that the approach can be extended beyond heralded state generation to the design of heralded gates.

Our approach allows determining the limitations of optical setups and measurement schemes, providing important insights into the resource requirements for quantum optical state generation and manipulation. In the next steps that the community is taking towards universal linear-optical quantum computing, some of the key milestones are measurement, heralding, and feed-forward. We hope that the present work will contribute to the discussion of what can and cannot be done.

We provide a MATLAB code of the NulLA applied to heralded optical state generation on GitHub \cite{OurCode}.

\section*{Acknowledgements}

This material is based upon work supported by the Air Force Office of Scientific Research under award number FA2386-23-1-4086. N.T.\ acknowledges support from the Australian Research Council through an Australian Research Council Discovery Early Career Researcher Award (DE220101082) and by the Alexander von Humboldt Foundation. This work was partially supported by the Australian Research Council Centre of Excellence for Quantum Computation and Communication Technology (Grant No.~CE170100012). 
J.E.~has been supported by the BMFTR  (QPIC-1, PhoQuant), the Cluster of Excellence ML4Q,
the DFG (SPP 2514),
Berlin Quantum, and the European Research 
Council
(DebuQC).

\vspace{1cm}
\appendix

\section{The use of equivalence classes}
\label{app:equivalenceclasses}
Ref.~\cite{PhysRevA.76.063808} proposed using equivalence classes for the transformation matrices to reduce the number of unknowns when setting up and solving the system of polynomial equations. This is achieved by recognizing and exploiting the fact that once a valid transformation matrix is found, other valid transformation matrices can be generated by scalar multiplications of certain rows and columns. Ref.~\cite{PhysRevA.76.063808} thus proposed using this freedom to set the corresponding elements of the first row and column of $A$ equal to one when constructing and solving the system of polynomial equations. Whilst the reduction of the number of unknowns in the system of polynomial equations is indeed highly desirable, this simplification comes with some unintended consequences, namely, a loss of generality. Specifically, setting an element equal to one is incompatible with a solution for $A$ where that element is zero.

As a very simple example, take the trivial task of generating the target state vector $\vert\psi_{ \mathrm{tar}} \rangle=|0,1\rangle$ from the input state vector $\vert\psi_{ \mathrm{in}} \rangle=|1,0\rangle$; this is deterministically achieved by the permutation transformation 
\begin{equation}
A=\left[\begin{matrix} 0 & 1 \\
1 & 0  \end{matrix}
\right],
\end{equation}
which implements a mode swap.
Using equivalence classes as introduced in Ref.~\cite{PhysRevA.76.063808} would set $A_{1,1}=1$, and because of this choice, the task would incorrectly be found to be infeasible when it is in fact feasible. 

Since we want to use the NulLA to establish no-go theorems, such incorrect assignment of infeasibility would be a serious problem. Therefore, we do not make use of equivalence classes in this work. However, if one were able to make use of equivalence classes by somehow ensuring such a situation is avoided, a significant speed-up could be achieved.\\

\section{Scaling details}
\label{app:scaling}

In this appendix, we will use the fact that the number of monomials in $V$ variables with degree up to $D$ is ${D+V \choose D}={D+V \choose V}$, and the number of monomials in $V$ variables with exactly degree $D$ is ${D+V-1\choose D}$.

  \emph{Number of polynomial equations}: The number of polynomial equations is the number of degree $(n-m)$ monomials possible in $(N-M)$ variables, where the degree stems from the number of photons left after heralding and the variables correspond to the creation operators in the $(N-M)$ target modes. Thus, the number of polynomial equations is ${n-m+N-M-1 \choose n-m}$. 

\emph{Degrees and number of variables in system of polynomial equations}: 
Each polynomial equation may contain one constant, and additionally contains monomials of degree $n$ in no more than $V_{\mathrm{max}}:=\mathrm{min}(Nn,N^2)$ variables, where the possible variables correspond to the $\mathrm{min}(Nn,N^2)$ elements of the $N\times N$ transformation matrix $A$ that can come into play if the $n$ input photons are in as many different input modes as possible. 

\emph{Number of unknowns in the linear system}: 
The number of unknowns in the linear system of the NulLA algorithm is upper bounded by ${n-m+N-M-1 \choose n-m}{V_{\mathrm{max}} + d \choose d}$; here, the first factor is the number of polynomial equations, because each polynomial equation gets its own set of unknowns. The second factor is the number of monomials in $V_{\mathrm{max}}=\mathrm{min}(Nn,N^2)$ variables (the variables of the system of polynomial equations) with degree up to $d$ (the infeasibility certificate degree).

\emph{Number of equations in the linear system}: The number of equations in the 
linear system of the NulLA algorithm is the number of different monomials that exist in the $\beta_i f_i$. 
The upper bound of this depends on whether $d<n-1$: 
\begin{itemize}
\item If $d\ge n-1$, the number of equations in the linear NulLA system is upper bounded by the number of monomials in $V_{\mathrm{max}}= \mathrm{min}(Nn,N^2)$ variables with degrees up to $(d+n)$, which is $V_{\mathrm{max}}+d+n \choose V_{\mathrm{max}}$.
\item If $d< n-1$, we can use our knowledge of the monomials that can exist in each polynomial equation (a constant and monomials of degree $n$), to further limit the upper bound, since the $\beta_i f_i$ do not contain any monomials with degrees between $d+1$ and $n-1$.
The number of equations in the linear system is thus upper bounded by ${V_{\mathrm{max}}+d+n \choose V_{\mathrm{max}}}-{V_{\mathrm{max}}-1+n \choose V_{\mathrm{max}}}+{V_{\mathrm{max}}+d \choose V_{\mathrm{max}}}$.
\end{itemize}

\bibliography{references}

\end{document}